\documentclass[12pt]{article}
\usepackage{cite}
\input epsf

\newcommand{\sect}[1]{\setcounter{equation}{0}\section{#1}}

\begin{document}

\title{A Covariant Entropy Conjecture}

\author{{\sc Raphael Bousso}\thanks{\it
    bousso@stanford.edu} \\[.3 ex]
    {\it Department of Physics, Stanford University,}\\
    {\it Stanford, California 94305-4060}
}

\date{SU-ITP-99-23~~~~~24 May 1999~~~~~hep-th/9905177}

\maketitle

\begin{abstract}

We conjecture the following entropy bound to be valid in all
space-times admitted by Einstein's equation: Let $A$ be the area of
any two-dimensional surface.  Let $L$ be a hypersurface generated by
surface-orthogonal null geodesics with non-positive expansion.  Let
$S$ be the entropy on $L$.  Then $S \leq A/4$.

We present evidence that the bound can be saturated, but not exceeded,
in cosmological solutions and in the interior of black holes.  For
systems with limited self-gravity it reduces to Bekenstein's bound.
Because the conjecture is manifestly time reversal invariant, its
origin cannot be thermodynamic, but must be statistical.  It thus
places a fundamental limit on the number of degrees of freedom in
nature.

\end{abstract}

\pagebreak

\sect{Introduction}
\label{sec-intro}

\subsection{Bekenstein's bound}
\label{sec-bekbound}

Bekenstein~\cite{Bek81} has proposed the existence of a universal
bound on the entropy $S$ of any thermodynamic system of total energy
$M$:
\begin{equation}
S \leq 2 \pi R M.
\label{eq-bek-em}
\end{equation}
$R$ is defined as the {\em circumferential
radius},
\begin{equation}
R = \sqrt{\frac{A}{4\pi}},
\end{equation}
where $A$ is the area of the smallest sphere circumscribing the
system.

For a system contained in a spherical volume, gravitational stability
requires that $M \leq R/2$.  Thus Eq.~(\ref{eq-bek-em}) implies
\begin{equation}
S \leq \frac{A}{4}.
\label{eq-bek-ea}
\end{equation}

The derivation of Eq.~(\ref{eq-bek-em}) involves a gedankenexperiment
in which the thermodynamic system is dropped into a Schwarzschild
black hole of much larger size.  The generalized second law of
thermodynamics~\cite{Bek72,Bek73,Bek74,Haw74} requires that the
entropy of the system should not exceed the entropy of the radiation
emitted by the black hole while relaxing to its original size; this
radiation entropy can be estimated~\cite{Bek94b,Pag76}.  Independently
of this fundamental derivation, the bound has explicitly been shown to
hold in wide classes of equilibrium systems~\cite{SchBek89}.

Bekenstein specified conditions for the validity of these bounds.  The
system must be of constant, finite size and must have limited
self-gravity, i.e., gravity must not be the dominant force in the
system.  This excludes, for example, gravitationally collapsing
objects, and sufficiently large regions of cosmological space-times.
Another important condition is that no matter components with negative
energy density are available.  This is because the bound relies on the
gravitational collapse of systems with excessive entropy, and is
intimately connected with the idea that information requires energy.
With matter of negative energy at hand, one could add entropy to a
system without increasing the mass, by adding enthropic matter with
positive mass as well as an appropriate amount of negative mass.  A
thermodynamic system that satisfies the conditions for the application
of Bekenstein's bounds will be called a {\em Bekenstein system}.

When the conditions set forth by Bekenstein are not satisfied, the
bounds can easily be violated.  The simplest example is a system
undergoing gravitational collapse.  Before it is destroyed on the
black hole singularity, its surface area becomes arbitrarily small.
Since the entropy cannot decrease, the bound is violated.  Or consider
a homogeneous spacelike hypersurface in a flat
Friedmann-Robertson-Walker universe.  The entropy of a sufficiently
large spherical volume will exceed the boundary area~\cite{FisSus98}.
This is because space is infinite, the entropy density is constant,
and volume grows faster than area.  From the point of view of
semi-classical gravity and thermodynamics, there is no reason to
expect that any entropy bound applies to such systems.

\subsection{Outline}
\label{sec-outline}

Motivated by the holographic conjecture~\cite{Tho93,Sus95}, Fischler
and Susskind~\cite{FisSus98} have suggested that some kind of entropy
bound should hold even for large regions of cosmological solutions,
for which Bekenstein's conditions are not satisfied.  However, no
fully general proposal has yet been formulated.  The Fischler-Susskind
bound~\cite{FisSus98}, for example, applies to universes which are not
closed or recollapsing, while other
prescriptions~\cite{EasLow99,Ven99,BakRey99,KalLin99,Bru99} can be
used for sufficiently small surfaces in a wide class of cosmological
solutions.

Our attempt to present a general proposal as arising, in a sense, from
fundamental considerations, should not obscure the immense debt we owe
to the work of others.  The importance of Bekenstein's seminal
paper~\cite{Bek81} will be obvious.  The proposal of Fischler and
Susskind~\cite{FisSus98}, whose influence on our prescription is
pervasive, uses light-like hypersurfaces to relate entropy and area.
This idea can be traced to the use of light-rays for formulating the
holographic principle~\cite{Tho93,Sus95,CorJac96}.  Indeed, Corley and
Jacobson~\cite{CorJac96} were the first to take a space-time (rather
than a static) point of view in locating the entropy related to an
area.  They introduced the concept of ``past and future screen-maps''
and suggested to choose only one of the two in different regions of
cosmological solutions.  Moreover, they recognized the importance of
caustics of the light-rays leaving a surface.  A number of authors
have investigated the application of Bekenstein's bound to
sufficiently small regions of the
universe~\cite{FisSus98,EasLow99,Ven99,BakRey99,KalLin99,Bru99}, and
have carefully exposed the difficulties that arise when such rules are
pushed beyond their range of
validity~\cite{FisSus98,EasLow99,KalLin99}.  These insights are
invaluable in the search for a general prescription.

We shall take the following approach.  We shall make no assumptions
involving holography anywhere in this paper.  Instead, we shall work
only within the framework of general relativity.  Taking Bekenstein's
bound as a starting point, we are guided mainly by the requirement of
covariance to a completely general entropy bound, which we conjecture
to be valid in arbitrary space-times (Sec.~\ref{sec-cec}).  Like
Fischler and Susskind~\cite{FisSus98}, we consider entropy not on
spacelike regions, but on light-like hypersurfaces.  There are four
such hypersurfaces for any given surface $B$.  We select portions of
at least two of them for an entropy/area comparison, by a covariant
rule that requires non-positive expansion of the generating null
geodesics.

In Sec.~\ref{sec-discussion}, we provide a detailed discussion of the
conjecture.  We translate the technical formulation into a set of
rules (Sec.~\ref{sec-recipe}).  In Sec.~\ref{sec-caustics}, we discuss
the criterion of non-positive expansion and find it to be quite
powerful.  We give some examples of its mechanism in
Sec.~\ref{sec-twoballs} and discover an effect that protects the bound
in gravitationally collapsing systems of high entropy.  In
Sec.~\ref{sec-spt} we establish a theorem that states the conditions
under which spacelike hypersurfaces may be used for entropy/area
comparisons.  Through this theorem, we recover Bekenstein's bound as a
special case.

In Sec.~\ref{sec-cosmology} we use cosmological solutions to test the
conjecture.  Our prescription naturally selects the apparent horizon
as a special surface.  In regions outside the apparent horizon
(Sec.~\ref{sec-antitrapped}), our bound is satisfied for the same
reasons that justified the Fischler-Susskind proposal within its range
of validity.  We also explain why reheating after inflation does not
endanger the bound.  For surfaces on or within the apparent horizon
(Sec.~\ref{sec-normal}), we find under worst case assumptions that the
covariant bound can be saturated, but not exceeded.  By requiring the
consistency of Bekenstein's conditions, we show that the apparent
horizon is indeed the largest surface whose interior one can hope to
treat as as a Bekenstein system.  This conclusion is later reached
independently, from the point of view of the covariant conjecture, in
Sec.~\ref{sec-coscol}.

In Sec.~\ref{sec-proposals} we discuss a number of recently proposed
cosmological entropy bounds.  Because of its symmetric treatment of
the four light-like directions orthogonal to any surface, which marks
its only significant difference from the Fischler-Susskind
bound~\cite{FisSus98}, our covariant prescription applies also to
closed and recollapsing universes (Sec.~\ref{sec-fs}).  Other useful
bounds~\cite{EasLow99,Ven99,BakRey99,KalLin99,Bru99} refer to the
entropy within a specified limited region.  The covariant bound can be
used to understand the range of validity of such bounds
(Sec.~\ref{sec-others}).  As an example, we apply a corollary derived
in Sec.~\ref{sec-coscol} to understand why the bound of
Ref.~\cite{BakRey99} cannot be applied to a flat universe with
negative cosmological constant~\cite{KalLin99}.

In Sec.~\ref{sec-collapse} we argue that the conjecture is true for
surfaces inside gravitationally collapsing objects.  We identify a
number of subtle mechanisms protecting the bound in such situations
(Sec.~\ref{sec-penetration}), and perform a quantitative test by
collapsing an arbitrarily large shell into a small region.  We find
again that the bound can be saturated but not exceeded.

In Sec.~\ref{sec-conclusions} we stress that the conjectured entropy
law is invariant under time reversal.  On the other hand, the physical
mechanisms responsible for its validity do not appear to be
T-invariant, and of course the very concept of thermodynamic entropy
has a built-in arrow of time.  Therefore the covariant bound must be
linked to the statistical origin of entropy.  Yet, it holds
independently of the microscopic properties of matter.  Thus the
covariant bound implies a fundamental limit on the total number of
independent degrees of freedom that are actually present in nature.
The holographic principle thus appears in this paper not as a
presupposition, but as a conclusion.

\paragraph{Notation and conventions}

We will ban formal definitions into footnotes, whenever they refer to
concepts which are intuitively clear.  We work with a manifold $M$ of
$3+1$ space-time dimensions, since the generalization to $D$
dimensions is obvious.  The terms {\em light-like} and {\em null} are
used interchangeably.  Any three-dimensional submanifold $H \subset M$
is called a {\em hypersurface} of $M$~\cite{HawEll}.  If two of its
dimensions are everywhere spacelike and the remaining dimension is
everywhere timelike (null, spacelike), $H$ will be called a {\em
timelike (null, spacelike) hypersurface}.  By a {\em surface} we
always refer to a two-dimensional spacelike submanifold $B \subset M$.
By a {\em light-ray} we never mean an actual electromagnetic wave or
photon, but simply a null geodesic.  We use the terms {\em congruence
of null geodesics}, {\em null congruence}, and {\em family of
light-rays} interchangeably.  A {\em light-sheet} of a surface $B$
will be defined in Sec.~\ref{sec-recipe} as a null hypersurface
bounded by $B$ and generated by a null congruence with non-increasing
expansion.  A number of definitions relating to Bekenstein's bound are
found on page~\pageref{bekdefs}.  We set $\hbar = c = G = k = 1$.

\pagebreak

\sect{The conjecture}
\label{sec-cec}

In constructing a covariant entropy bound, one first has to decide
whether to aim at an entropy/mass bound, as in Eq.~(\ref{eq-bek-em}),
or an entropy/area bound, as in Eq.~(\ref{eq-bek-ea}).  Local energy
is not well-defined in general relativity, and for global definitions
of mass the space-time must possess an infinity~\cite{Wald,MTW}.  This
eliminates any hope of obtaining a completely general bound involving
mass.  Area, on the other hand, can always be covariantly defined as
the proper area of a surface.

Having decided to search for an entropy/area bound, our difficulty
lies not in the quantitative formula, $S \leq A/4$; this will remain
unchanged.  The problem we need to address is the following: Given a
two-dimensional surface $B$ of area $A$, on which hypersurface $H$
should we evaluate the entropy $S$?  We shall retain the rule that $B$
must be a boundary of $H$.  In general space-times, this leaves an
infinite choice of different hypersurfaces.  Below we will construct
the rule for the hypersurfaces, guided by a demand for symmetry and
consistency with general relativity.

As a starting point, we write down a slightly generalized version of
Bekenstein's entropy/area bound, Eq.~(\ref{eq-bek-ea}): {\em Let $A$
be the area of any closed two-dimensional surface $B$, and let $S$ be
the entropy on the spatial region $V$ enclosed by $B$.  Then $S \leq
A/4$.}

Bekenstein's bounds were derived for systems of limited self-gravity
and finite extent (finite spatial region $V$).  In order to be able to
implement general coordinate invariance, we shall now drop these
conditions.  Of course this could go hopelessly wrong.  The
generalized second law of thermodynamics gives no indication of any
useful entropy bound if Bekenstein's conditions are not met.  Ignoring
such worries, we move on to ask how the formulation of the bound has
to be modified in order to achieve covariance.

An obvious problem is the reference to ``the'' spatial region.  If we
demand covariance, there cannot be a preferred spatial hypersurface.
Either the bound has to be true for {\em any} spatial hypersurface
enclosed by $B$, or we have to insist on light-like hypersurfaces.
But the possibility of using {\em any} spatial hypersurface is already
excluded by the counterexamples given at the end of
Sec.~\ref{sec-bekbound}; this was first pointed out in
Ref.~\cite{FisSus98}.

\begin{figure}[htb!]
  \hspace{.25\textwidth} \vbox{\epsfxsize=.5\textwidth
  \epsfbox{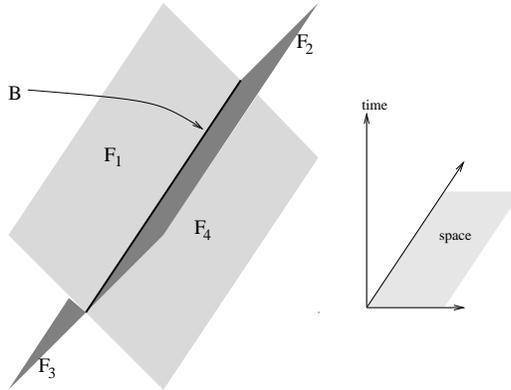}}
\caption%
{\small\sl There are four families of light-rays projecting
orthogonally away from a two-dimensional surface $B$, two
future-directed families (one to each side of $B$) and two
past-directed families.  At least two of them will have non-positive
expansion.  The null hypersurfaces generated by non-expanding
light-rays will be called ``light-sheets.''  The covariant entropy
conjecture states that the entropy on any light-sheet cannot exceed a
quarter of the area of $B$.}
\label{fig-foursurfaces}
\end{figure}

Therefore {\em we must use null hypersurfaces bounded by $B$}.  The
natural way to construct such hypersurfaces is to start at the surface
$B$ and to follow a family of light-rays (technically, a ``congruence
of null geodesics'') projecting out orthogonally%
\footnote{While it may be clear what we mean by light-rays which are
orthogonal to a closed surface $B$, we should also provide a formal
definition.  In a convex normal neighbourhood of $B$, the boundary of
the chronological future of $B$ consists of two future-directed null
hypersurfaces, one on either side of $B$ (see Chapter 8 of
Wald~\cite{Wald} for details).  Similarly, the boundary of the
chronological past of $B$ consists of two past-directed null
hypersurfaces.  Each of these four null hypersurfaces is generated by
a congruence of null geodesics starting at $B$.  At each point on $p
\in B$, the four {\em null directions orthogonal to $B$} are defined
by the tangent vectors of the four congruences.  This definition can
be extended to smooth surfaces $B$ with a boundary $\partial B$: For
$p \in \partial B$, the four orthogonal null directions are the same
as for a nearby point $q \in B - \partial B$, in the limit of
vanishing proper distance between $p$ and $q$.  We will also allow $B$
to be on the boundary of the space-time $M$, in which case there will
be fewer than four options.  For example, if $B$ lies on a boundary of
space, only the ingoing light-rays will exist.  We will not make such
exceptions explicit in the text, as they are obvious.}
from $B$.  But we have four choices: the family of light-rays can be
future-directed outgoing, future-directed ingoing, past-directed
outgoing, and past-directed ingoing (see Fig.~\ref{fig-foursurfaces}).
Which should we select?  And how far may we follow the light-rays?

In order to construct a selection rule, let us briefly return to the
limit in which Bekenstein's bound applies.  For a spherical surface
around a Bekenstein system, the enclosed entropy cannot be larger than
the area.  But the same surface is also a boundary of the infinite
region on its outside.  The entropy outside could clearly be anything.
From this we learn that it is important to {\em consider the entropy
only on hypersurfaces which are not outside the boundary}.

But what does ``outside'' mean in general situations?  The side that
includes infinity?  Then what if space is closed?  Fortunately, there
exists an intuitive notion of ``inside'' and ``outside'' that is
suitable to be generalized to a covariant rule.  Think of ordinary
Euclidean geometry, and start from a closed surface $B$.  Construct a
second surface by moving every point on $B$ an infinitesimal distance
away to one side of $B$, along lines orthogonal to $B$.  If this
increases the area, then we say that we have moved outside.  If the
area decreases, we have moved inside.

This consideration yields the selection rule.  {\em We start at $B$,
and follow one of the four families of orthogonal light-rays, as long
as the cross-sectional area is decreasing or constant.  When it
becomes increasing, we must stop.}  This can be formulated technically
by demanding that the expansion%
\footnote{The expansion $\theta$ of a congruence of null geodesics is
defined, e.g., in Refs.~\cite{HawEll,Wald} and will be discussed
further in Sec.~\ref{sec-caustics}.  It measures the local rate of
change of the cross-sectional area as one moves along the
light-rays. Let $\lambda$ be an affine parameter along a family of
light-rays orthogonal to $B$.  Let ${\cal A}(\lambda)$ denote the area
of the surfaces spanned by the light-rays at the affine time
$\lambda$.  Then ${\cal A}(0) = A$, the area of $B$.  $\cal
A(\lambda)$ is independent of the choice of Lorentz frame~\cite{MTW},
so it is a covariant quantity.  Then $\theta = \frac{d {\cal A} / d
\lambda}{{\cal A}}$.  We choose the affine parameter to be increasing
in the directions away from the surface $B$ (this implies that we are
using a different affine parameter $\lambda_i$ for each of the four
null-congruences).  Non-positive expansion, $\theta \leq 0$, thus
means that the cross-sectional area is not increasing in the direction
away from $B$.}
of the orthogonal null congruence must be non-positive, in the
direction away from the surface $B$.  By continuity across $B$, the
expansion of past-directed light-rays going to one side is the
negative of the expansion of future-directed light-rays heading the
other way.  Therefore we can be sure that at least two of the four
null directions will be allowed.  If the expansion of one pair of null
congruences vanishes on $B$, there will be three allowed directions.
If the expansion of both pairs vanishes on $B$, all four null
directions will be allowed.

This covariant definition of ``inside'' and ``outside'' does not
require the surface $B$ to be closed.  Only the naive definition, by
which ``inside'' was understood to mean a finite region delimited by a
surface, needed the surface to be closed.  Therefore we shall now drop
this condition and allow any connected two-dimensional surface.  This
enables us to assume without loss of generality that the expansion in
each of the four null directions does not change sign anywhere on the
surface $B$.  If it does, we simply split $B$ into suitable domains
and apply the entropy law to each domain individually.

Finally, in an attempt to protect our conjecture against pathologies
such as superluminal entropy flow, we will require the dominant energy
condition to hold: for all timelike $v_a$, $T^{ab} v_a v_b \geq 0$ and
$T^{ab} v_a$ is a non-spacelike vector.  This condition states that to
any observer the local energy density appears non-negative and the
speed of energy flow of matter is less than the speed of light.  It
implies that a space-time must remain empty if it is empty at one time
and no matter is coming in from infinity~\cite{HawEll}.  It is
believed that all physically reasonable forms of matter satisfy the
dominant energy condition~\cite{HawEll,Wald}, so we are not imposing a
significant restriction.  It may well be possible, however, to weaken
the assumption further; this is under investigation%
\footnote{We do not wish to permit matter with negative energy
density, since that would open the possibility of creating arbitrary
amounts of positive energy matter, and thus arbitrary entropy, by
simultaneously creating negative energy matter.  Worse still, if such
a process can be reversed, one would be able to destroy entropy and
violate the second law.  Therefore, negative energy matter permitting
such processes must not be allowed in a physical theory.  A negative
cosmological constant is special in that it cannot be used for such a
process.  Indeed, we are currently unaware of any counterexamples to
the covariant bound in space-times with a negative cosmological
constant (see Secs.~\ref{sec-spt} and~\ref{sec-others}).  This
suggests that it may be sufficient to demand only causal energy flow
(for all timelike $v_a$, $T^{ab} v_a$ is a non-spacelike vector),
without requiring the positivity of energy. --- We wish to thank
Nemanja Kaloper and Andrei Linde for a discussion of these issues.}.

We should also require that the space-time is inextendible and
contains no null or timelike (``naked'') singularities.  This is
necessary if we wish to exclude the possibility of destroying or
creating arbitrary amounts of entropy on such boundaries.  These
conditions are believed to hold in any physical space-time, so we
shall not spell them out below.  (Of course we are not excluding the
spacelike singularities occuring in cosmology and in gravitational
collapse; indeed, much of this paper will be devoted to investigating
the validity of the proposed bound in the vicinity of such
singularities.)  We thus arrive at a conjecture for a covariant bound
on the entropy in any space-time.

\paragraph{Covariant Entropy Conjecture}
{\em Let $M$ be a four-dimensional space-time on which Einstein's
equation is satisfied with the dominant energy condition holding for
matter.  Let $A$ be the area of a connected two-dimensional spatial
surface $B$ contained in $M$.  Let $L$ be a hypersurface bounded by
$B$ and generated by one of the four null congruences orthogonal to
$B$.  Let $S$ be the total entropy contained on $L$.  If the expansion
of the congruence is non-positive (measured in the direction away from
$B$) at every point on $L$, then $S \leq A/4$.}

\sect{Discussion}
\label{sec-discussion}

\subsection{The recipe}
\label{sec-recipe}

In the previous paragraph, we formulated the conjecture with an eye on
formal accuracy and generality.  For all practical purposes, however,
it is more useful to translate it into a set of rules like the
following recipe (see Fig.~\ref{fig-foursurfaces}):

\begin{enumerate}

\item{Pick any two-dimensional surface $B$ in the space-time $M$.}

\item{There will be four families of light-rays projecting orthogonally
away from $B$ (unless $B$ is on a boundary of $M$): $F_1 \ldots F_4$.}

\item{As shown in the previous section, we can assume without loss of
generality that the expansion of $F_1$ has the same sign everywhere on
$B$.  If the expansion is positive (in the direction away from $B$),
i.e., if the cross-sectional area is increasing, $F_1$ must not be
used for an entropy/area comparison.  If the expansion is zero or
negative, $F_1$ will be allowed.  Repeating this test for each family,
one will be left with at least two allowed families.  If the expansion
is zero in some directions, there may be as many as three or four
allowed families.}

\item{Pick one of the allowed families, $F_i$.  Construct a null
hypersurface $L_i$, by following each light-ray until one of the
following happens:

  \begin{enumerate}

  \item{The light-ray reaches a boundary or a singularity of the
  space-time.}

  \item{The expansion becomes positive, i.e., the cross-sectional area
  spanned by the family begins to increase in a neighborhood of the
  light-ray.}

  \end{enumerate}

The hypersurface $L_i$ obtained by this procedure will be called a
{\em light-sheet} of the surface $B$.  For every allowed family, there
will be a different light-sheet.}

\item{The conjecture states that the entropy $S_i$ on the light-sheet
$L_i$ will not exceed a quarter of the area of $B$:
\begin{equation}
S_i \leq \frac{A}{4}.
\end{equation}
Note that the bound applies to each light-sheet individually.  Since
$B$ may possess up to four light-sheets, the total entropy on all
light-sheets could add up to as much as $A$.}

\end{enumerate}

We should add some remarks on points 2 and 3.  In many situations it
will be natural to call $F_1 \ldots F_4$ the future-directed ingoing,
future-directed outgoing, past-directed ingoing and past-directed
outgoing family of surface-orthogonal geodesics.  But we stress that
``ingoing'' and ``outgoing'' are just arbitrary labels distinguishing
the two sides of $B$; only if $B$ is closed, there might be a
preferred way to assign these names.  Our rule does not refer to
``ingoing'' and ``outgoing'' explicitly.

In fact the covariant entropy bound does not even refer to ``future''
and ``past.''  The conjecture is manifestly time reversal invariant.
We regard this as its most significant property.  After all,
thermodynamic entropy is never T-invariant, and neither is the
generalized second law of thermodynamics, which underlies Bekenstein's
bound.  This will be discussed further in Sec.~\ref{sec-spt}.  One can
draw strong conclusions from these simple observations
(Sec.~\ref{sec-conclusions}).

If $B$ is a closed surface, we can characterize it as {\em trapped},
{\em anti-trapped} or {\em normal} (see Refs.~\cite{HawEll,Wald} for
definitions).  This provides a simple criterion for the allowed
families.  If $B$ is trapped (anti-trapped), it has two
future-directed (past-directed) light-sheets.  If $B$ is normal, it
has a future-directed and a past-directed light-sheet on the same
side, which is usually called the inside.  If $B$ lies on an {\em
apparent horizon} (the boundary between a trapped or anti-trapped
region and a normal region), it can have more than two light-sheets.
For example, if $B$ is {\em marginally outer trapped}~\cite{HawEll},
i.e., if the future-directed outgoing geodesics have zero convergence,
then it has two future-directed light-sheets and a past-directed
ingoing light-sheet.

\subsection{Caustics as light-sheet endpoints}
\label{sec-caustics}

We have defined a light-sheet to be a certain subset of the null
hypersurface generated by an ``allowed'' family of light-rays.  The
rule is to start at the surface $B$ and to follow the light-rays only
as long as the expansion is zero or negative.  In order to understand
if and why the conjecture may be true, we must understand perfectly
well what it is that can cause the expansion to become positive.
After all, it is this condition alone which must prevent the
light-sheet from sampling too much entropy and violating the
conjectured bound.  Our conclusion will be that the expansion becomes
positive only at caustics. The simplest example of such a point is the
center of a sphere in Minkowski space, at which all ingoing light-rays
intersect.

Raychauduri's equation for a congruence of
null geodesics with tangent vector field $k^a$ and affine parameter
$\lambda$ is given by
\begin{equation}
\frac{d\theta}{d\lambda} = - \frac{1}{2} \theta^2
  - \hat{\sigma}_{ab} \hat{\sigma}^{ab}
  - 8 \pi T_{ab} k^a k^b
  + \hat{\omega}_{ab} \hat{\omega}^{ab},
\label{eq-raych}
\end{equation}
where $T_{ab}$ is the stress-energy tensor of matter.  The expansion
$\theta$ measures the local rate of change of an element of
cross-sectional area ${\cal A}$ spanned by nearby geodesics:
\begin{equation}
\theta = \frac{1}{\cal A} \frac{d \cal A}{d\lambda}.
\end{equation}
The vorticity $\hat{\omega}_{ab}$ and shear $\hat{\sigma}_{ab}$ are
defined in Refs.~\cite{HawEll,Wald}.  The vorticity vanishes for
surface-orthogonal null congruences.  The first and second term on the
right hand side are manifestly non-positive.  The third term will be
non-positive if the {\em null convergence condition}~\cite{HawEll}
holds:
\begin{equation}
T_{ab} k^a k^b \geq 0~~\mbox{for all null}~k^a.
\label{eq-nullconv}
\end{equation}
The dominant energy condition, which we are assuming, implies that the
null convergence condition will hold.  (It is also implied by the weak
energy condition, or by the strong energy condition.)

Therefore the right hand side of Eq.~(\ref{eq-raych}) is non-positive.
It follows that $\theta$ cannot increase along any geodesic.  (This
statement is self-consistent, since the sign of $\theta$ changes if we
follow the geodesic in the opposite direction.)  Then how can the
expansion ever become positive?  By dropping two of the non-positive
terms in Eq.~(\ref{eq-raych}), one obtains the inequality
\begin{equation}
\frac{d\theta}{d\lambda}  \leq  - \frac{1}{2} \theta^2.
\label{eq-raych2}
\end{equation}
If the expansion takes the negative value $\theta_0$ at any point on a
geodesic in the congruence, Eq.~(\ref{eq-raych2}) implies that the
expansion will become negative infinite, $\theta \rightarrow -\infty$,
along that geodesic within affine time $\Delta \lambda \leq
2/|\theta_0|$~\cite{Wald}.  This can be interpreted as a {\em
caustic}.  Nearby geodesics are converging to a single focal point,
where the cross-sectional area $\cal A$ vanishes.  When they
re-emerge, the cross-sectional area starts to increase.  Thus, the
expansion $\theta$ jumps from $-\infty$ to $+\infty$ at a caustic.
Then the expansion is positive, and we must stop following the
light-ray.  This is why caustics form the endpoints of the light-sheet.

\subsection{Light-sheet examples and first evidence}
\label{sec-twoballs}

The considerations in Sec.~\ref{sec-caustics} allow us to rephrase the
rule for constructing light-sheets: Follow each light-ray in an
allowed family until a caustic is reached.  The effect of this
prescription can be understood by thinking about closed surfaces in
Minkowski space.  The simplest example is a spherical surface.  The
past-directed outgoing and future-directed outgoing families are
forbidden, because they have positive expansion.  The past- and
future-directed ingoing families are allowed.  Both encounter caustics
when they reach the center of the sphere.  Thus they each sweep the
interior of the sphere exactly once.  If we deform the surface into a
more irregular shape, such as an ellipsoid, there may be a line, or
even a surface, of caustics at which the light-sheet ends.  In some
cases (e.g. if the surface is a box and does not enclose much matter),
non-neighbouring light-rays may cross.  This does not constitute a
caustic,%
\footnote{We thank Ted Jacobson for pointing this out.}
and the light-rays need not be terminated there.  They can go on until
they are bent into caustics by the matter they encounter.  Thus some
of the entropy may be counted more than once.  This is merely a
consequence of a desirable feature of our prescription: that it is
local and applies separately to every infinitesimal part of any
surface.

\begin{figure}[htb!]
  \hspace{.05\textwidth} \vbox{\epsfxsize=.9\textwidth
  \epsfbox{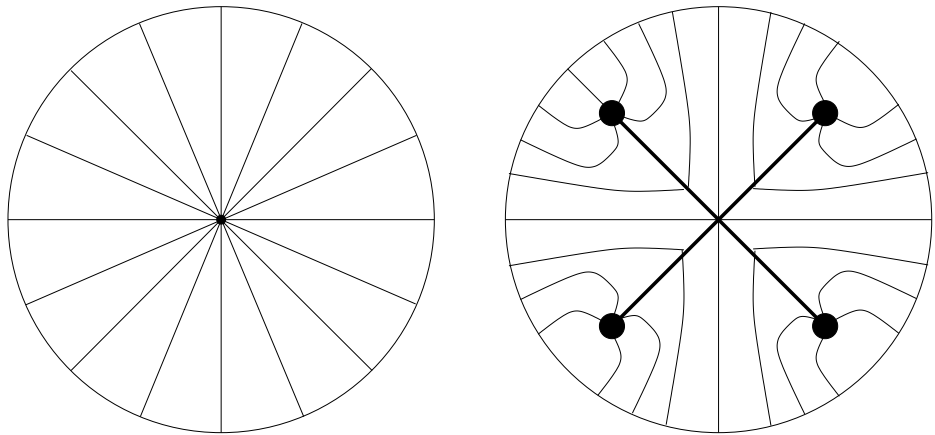}}
\caption%
{\small\sl Light-rays from a spherical surface into a massive body.
If the body is spherically symmetric internally (left), the light-rays
all meet at a ``radial'' caustic in the center of the sphere.  If the
internal mass distribution is not spherically symmetric, most rays get
deflected into angular directions and end on ``angular caustics.''
The picture on the right shows an example with four overdense regions.
The thick lines represent caustics.}
\label{fig-angcaustics}
\end{figure}

For a spherical surface surrounding a spherically symmetric body of
matter, the ingoing light-rays will end on a caustic in the center, as
for an empty sphere.  If the interior mass distribution is not
spherically symmetric, however, some light-rays will be deflected into
angular directions, and will form ``angular caustics'' (see
Fig.~\ref{fig-angcaustics}).
This does not mean that the interior will not be completely swept out
by the light-sheet.  Between two light-rays that get deflected into
different overdense regions, there are infinitely many light-rays that
proceed further inward.  It does mean, however, that we have to follow
some of the light-rays for a much longer affine time than we would in
the spherically symmetric case.

\begin{figure}[htb!]
  \hspace{.05\textwidth} \vbox{\epsfxsize=.9\textwidth
  \epsfbox{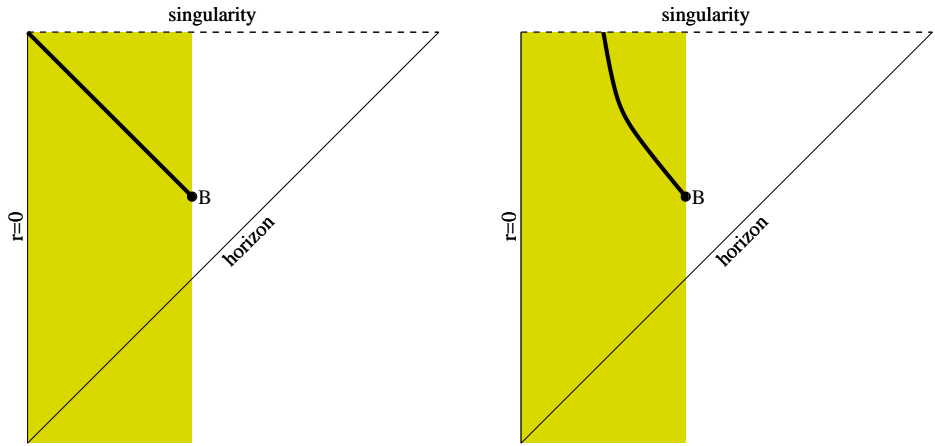}}
\caption%
{\small\sl Two Penrose diagrams for collapsing spherical objects of
identical mass.  Each point represents a two-sphere.  The thick lines
are light-rays.  In an object which is spherically symmetric
internally (left), the future-directed ingoing light-rays originating
from its surface $B$ (at a suitable time) just make it to the center
of the object before reaching the singularity.  In a highly disordered
system, on the other hand, light-rays will be deflected into angular
directions (right).  We are taking the surface $B$ to have the same
area as before.  The light-ray path looks timelike because the
angular directions are suppressed in the diagram.  We see that
the light-rays will not sweep the entire interior of the
disordered system before meeting the singularity.}
\label{fig-twoballs}
\end{figure}

This does not make a difference for static systems: they will be
completely penetrated by the light-sheet in any case.  In a system
undergoing gravitational collapse, however, light-rays will hit the
future singularity after a finite affine time.  Consider a collapsing
ball which is exactly spherically symmetric, and a future-ingoing
light-sheet starting at the outer surface of the system, when it is
already within its own Schwarzschild horizon.  We can arrange things
so that the light-sheet reaches the caustic at $r=0$ exactly when it
also meets the singularity (see Fig.~\ref{fig-twoballs}).
Now consider a different collapsing ball, of identical mass, and
identical radius when the light-rays commence.  While this ball may be
spherically symmetric on the large scale, let us assume that it is
highly disordered internally.  The light-rays will thus be deflected
into angular directions.  As Fig.~\ref{fig-angcaustics} illustrates,
this means that they take intricate, long-winded paths through the
interior: they ``percolate.''  This consumes more affine time than the
direct path to the center taken in the first system.  Therefore the
second system will not be swept out completely before the singularity
is reached (see Fig.~\ref{fig-twoballs}).

Why did we spend so much time on this example?  The first ball is a
system with low entropy, while the second ball has high entropy.  One
might think that no kind of entropy bound can apply when a highly
enthropic system collapses: the surface area goes to zero, but the
entropy cannot decrease.  The above considerations have shown,
however, that light-sheets percolate rather slowly through highly
enthropic systems, because the geodesics follow a kind of random walk.
Since they end on the black hole singularity within finite affine
time, they sample a smaller portion of a highly enthropic system than
they would for a more regular system.  (In Fig.~\ref{fig-angcaustics},
the few light-rays that go straight to the center of the inhomogeneous
system also take more affine time to do so than in a homogeneous
system, because they pass through an underdense region.  In a
homogeneous system, they would encounter more mass; by Raychauduri's
equation, this would accelerate their collapse.)  Therefore it is in
fact quite plausible, if counter-intuitive, that the covariant entropy
bound holds even during the gravitational collapse of a system
initially saturating Bekenstein's bound.

In Sec.~\ref{sec-collapse}, we will discuss additional constraints on
the penetration depth of light-sheets in collapsing highly enthropic
systems, and a quantitative test will be performed.

\subsection{Recovering Bekenstein's bound}
\label{sec-spt}

The covariant entropy conjecture can only be sensible if we can
recover Bekenstein's bound from it in an appropriate limit.  For a
Bekenstein system (see Sec.~\ref{sec-bekbound}: a thermodynamic system
of constant, finite size and limited self-gravity), the boundary area
should bound the entropy {\em in the spatial region} occupied by the
system.  The covariant bound, on the other hand, uses null
hypersurfaces to compare entropy and area.  Then how can Bekenstein's
bound be recovered?

While null hypersurfaces are certainly required in general, it turns
out that there is a wide class of situations in which an entropy bound
on spacelike hypersurfaces can be inferred from the covariant entropy
conjecture.  We will now identify sufficient conditions and derive a
theorem on the entropy of spatial regions.  We will then show that
Bekenstein's entropy/area bound is indeed implied by the covariant
bound, namely as a special case of the theorem.

Let $A$ be the area of a closed surface $B$ possessing at least
one future-directed light-sheet $L$.  Suppose that $L$ has no boundary
other than $B$.  Then we shall call the direction of this light-sheet
the ``inside'' of $B$.  Let the spatial region $V$ be the interior of
$B$ on some spacelike hypersurface through $B$.  If the region $V$ is
contained in the causal past of the light-sheet $L$, the dominant
energy condition implies that all matter in the region $V$ must
eventually pass through the light-sheet $L$.  Then the second law of
thermodynamics implies that the entropy on $V$, $S_V$, cannot exceed
the entropy on $L$, $S_L$.  By the covariant entropy bound, $S_L \leq
A/4$.  It follows that the entropy of the spatial region $V$ cannot
exceed a quarter of its boundary area: $S_V \leq A/4$.

The condition that the future-directed light-sheet $L$ contain no
boundaries makes sure that none of the entropy of the spatial region
$V$ escapes through holes in $L$.  Neither can any of the entropy
escape into a black hole singularity, because we have required that
the spatial region must lie in the causal past of $L$.  Since we are
always assuming that the space-time is inextendible and that no naked
singularities are present, all entropy on $V$ must go through $L$.  We
summarize this argument in the following theorem.

\paragraph{Spacelike Projection Theorem}
{\em Let $A$ be the area of a closed surface $B$ possessing a
future-directed light-sheet $L$ with no boundary other than $B$.  Let
the spatial region $V$ be contained in the intersection of the causal
past of $L$ with any spacelike hypersurface containing $B$.  Let $S$
be the entropy on $V$.  Then $S \leq A/4$.}\\

Now consider, in asymptotically flat space, a Bekenstein system in a
spatial region $V$ bounded by a closed surface $B$ of area $A$.  The
future-directed ingoing light-sheet $L$ of $B$ exists (otherwise $B$
would not have ``limited self-gravity''), and can be taken to end
whenever two (not necessarily neighbouring) light-rays meet.  Thus it
will have no other boundary than $B$.  Since the gravitational binding
of a Bekenstein system is not strong enough to form a black hole, $V$
will be contained in the causal past of $L$.  Therefore the conditions
of the theorem are satisfied, and the entropy of the system must be
less than $A/4$.  We have recovered Bekenstein's bound.

There are many other interesting applications of the theorem.  In
particular, it can be used to show that area bounds entropy on
spacelike sections of anti-de~Sitter space.  This can be seen by
taking $B$ to be any sphere, at any given time.  The future-ingoing
light-sheet of $B$ exists, and unless the space contains a black hole,
has no other boundary.  Its causal past includes the entire space
enclosed by $B$.  This remains true for arbitrarily large spheres, and
in the limit as $B$ approaches the boundary at spatial infinity.

The theorem is immensely useful, because it essentially tells us under
which conditions we can treat a region of space as a Bekenstein
system.  In general, however, the light-sheets prescribed by the
covariant entropy bound provide the only consistent way of comparing
entropy and area.

We pointed out in Sec.~\ref{sec-recipe} that the covariant entropy
bound is T-invariant.  The spacelike projection theorem is not
T-invariant; it refers to past and future explicitly.  This is because
the second law of thermodynamics enters its derivation.  The asymmetry
is not surprising, since Bekenstein's bound, which we recovered by the
theorem, rests on the second law.  We should be surprised only when an
entropy law {\em is} T-invariant.  It is this property which forces us
to conclude that the origin of the covariant bound is not
thermodynamic, but statistical (Sec.~\ref{sec-conclusions}).

\pagebreak

\sect{Cosmological tests}
\label{sec-cosmology}

The simplest way to falsify the conjecture would be to show that it
conflicts directly with observation.  In this section, we will apply
the covariant bound to the cosmological models that our universe is
believed to be described by (and to many more by which it certainly is
not described).  Since we claim that the conjecture is a universal law
valid for all space-times satisfying Einstein's equations (with the
dominant energy condition holding for matter), it must be valid for
cosmological solutions in particular.  We will see that it passes the
test, and more significantly, it just passes it.

We consider Friedmann-Robertson-Walker (FRW) cosmologies, which are
described by a metric of the form
\begin{equation}
ds^2 = -dt^2 + a^2(t) \left( \frac{dr^2}{1-kr^2} + r^2 d\Omega^2
\right).
\label{eq-FRW1}
\end{equation}
An alternative form uses comoving coordinates:
\begin{equation}
ds^2 = a^2(\eta) \left[ -d\eta^2 + d\chi^2 + f^2(\chi) d\Omega^2
\right].
\label{eq-FRW2}
\end{equation}
Here $k = -1$, $0$, $1$ and $f(\chi) = \sinh \chi$, $\chi$, $\sin
\chi$ correspond to open, flat, and closed universes respectively.

The {\em Hubble horizon} is the inverse of the expansion rate $H$:
\begin{equation}
r_{\rm HH} = H^{-1} = \frac{a}{da/dt}.
\label{eq-hubble}
\end{equation}
The {\em particle horizon} is the distance travelled by light since
the big bang:
\begin{equation}
\chi_{\rm PH} = \eta.
\label{eq-particle}
\end{equation}
The {\em apparent horizon} is defined geometrically as a sphere at
which at least one pair of orthogonal null congruences have zero
expansion.  It is given by~\cite{BakRey99}
\begin{equation}
r_{\rm AH} = \frac{1}{\sqrt{H^2 + \frac{k}{a^2}}}.
\label{eq-ah-hubble}
\end{equation}
Using Friedmann's equation,
\begin{equation}
H^2 = \frac{8\pi \rho}{3} - \frac{k}{a^2},
\end{equation}
one finds
\begin{equation}
r_{\rm AH} = \sqrt{\frac{3}{8 \pi \rho}},
\label{eq-ah-rho}
\end{equation}
where $\rho$ is the energy density of matter.

We consider matter described by $T_{ab} = \mbox{diag}(\rho,p,p,p)$,
with pressure $p = \gamma \rho$.  The dominant energy condition
requires that $\rho \geq 0$ and $-1 \leq \gamma \leq 1$.  The case
$\gamma = -1$ is special, because it leads to a different global
structure from the other solutions.  It corresponds to de~Sitter
space, which has no past or future singularity.  This solution is
significant because it describes an inflationary universe.  We will
comment on inflation at the end of Sec.~\ref{sec-antitrapped}.

We will test the conjecture on spherical surfaces $B$ characterized by
some value of $r$, or of $(\eta,\chi)$.  As we found at the end of
Sec.~\ref{sec-recipe}, the directions of the light-sheets of a surface
depend on its classification as trapped, normal, or anti-trapped.  In
the vicinity of $\chi=0$ (and for closed universes, also on the
opposite pole, near $\chi=\pi$), the spherical surfaces will be
normal.  The larger spheres beyond the apparent horizon(s) will be
anti-trapped.  Some universes, for example most closed universes, or a
flat universe with negative cosmological constant~\cite{KalLin99},
recollapse.  Such universes necessarily contain trapped surfaces.  But
trapped regions can occur in any case by gravitational collapse.  The
surfaces in the interior of such regions provide a serious challenge
for the covariant entropy conjecture, because their area shrinks to
zero while the enclosed entropy cannot decrease.  We address this
problem in some generality in Sec.~\ref{sec-collapse}, where we argue
that the conjecture holds even in such regions.  Because of its
importance, the special case of the adiabatic recollapse of a closed
universe will be treated explicitly in Sec.~\ref{sec-fs}.  In the
present section, we shall therefore discuss only normal and
anti-trapped surfaces.

\subsection{Anti-trapped surfaces}
\label{sec-antitrapped}

An anti-trapped surface $B$ contains two past-directed light-sheets.
Unless $B$ lies within the particle horizon (of either pole in the
closed case), both light-sheets will be ``truncated'' at the Planck
era near the past singularity.  The truncation has the desirable
effect that the volume swept by the light-sheets grows not like
$A^{3/2}$, but roughly like the area.  In fact, the ``ingoing''
light-sheet coincides with the ``truncated lightcones'' used in the
entropy conjecture of Fischler and Susskind~\cite{FisSus98}, and the
other light-sheet can be treated similarly.  In open universes the
bound will be satisfied more comfortably than in flat
ones~\cite{FisSus98}.  The validity of the bound was checked for flat
universes with $0 \leq \gamma \leq 1$ in Ref.~\cite{FisSus98} and for
$-1 < \gamma < 0$ in Ref.~\cite{KalLin99}.  Here we will give a
summary of these results, and we will explain why reheating after
inflation does not violate the covariant bound.  Subtleties arising in
closed universes will be discussed separately in Sec.~\ref{sec-fs}.

The reason why the bound is satisfied near a past singularity is
simple.  The first moment of time that one can sensibly talk about is
one Planck time after the singularity.  At this time, there cannot
have been more than one unit of entropy per Planck volume, up to a
factor of order one.  This argument does not involve any assumptions
about ``holography;'' we are merely applying the usual Planck scale
cutoff.  We cannot continue light-sheets into regions where we have no
control over the physics.  A backward light-sheet of an area $A$
specified at $t=2 t_{\rm Pl}$ will be truncated at $t = t_{\rm Pl}$.
It will sweep a volume of order $A l_{\rm Pl}$ and the entropy bound
will at most be saturated.

Let us define $\sigma$ as the entropy/area ratio at the Planck time.
Consider a universe filled with any type of matter allowed by the
dominant energy condition, $-1 \leq \gamma \leq 1$.  We may exclude
$\gamma = -1$, since de Sitter space does not contain a singularity in
the past.  The scale factor will be given by $a(t) =
t^{\frac{2}{3(\gamma+1)}}$.  If the evolution is adiabatic, the ratio
of entropy to area behaves as~\cite{KalLin99}
\begin{equation}
\frac{S}{A} \leq \sigma t^{\frac{\gamma-1}{\gamma+1}}
\end{equation}
for any past-directed light-sheet of areas specified at a later time
$t$.  (Equality holds, e.g., for spherical areas at least as large as
the particle horizon.)  The exponent is non-positive, so the
entropy/area ratio does not increase.  Since the bound is satisfied at
the Planck time, it will remain satisfied later.

Another way to see this is to consider the particle horizon $n$
(e.g. $n=10$) Planck times after the big bang of a flat FRW universe.
Its area will be $O(n^2)$ and its past-directed light-sheet will sweep
$O(n^3)$ Planck volumes.  Let us assume that each of these Planck
volumes contains one unit of entropy; then the bound would be
violated.  But most of these volumes are met by the light-sheet at a
time later than $t_{\rm Pl}$.  Therefore there must have been, at the
Planck time, Planck volumes containing more than a unit of entropy,
which is impossible.  (Note that this argument would break down if we
allowed naked singularities!)

If the evolution is non-adiabatic, the entropy bound nevertheless
predicts that $S/A \leq 1/4$, implying that there is a limit on how
rapidly the universe can produce entropy.  This will be further
discussed in Sec.~\ref{sec-normal}.

\paragraph{Inflation}
Of course the notion that the standard cosmology extends all the way
back to the Planck era is not seriously tenable.  In order to
understand essential properties of our universe, such as its
homogeneity and flatness, and its perturbation spectrum, it is usually
assumed that the radiation dominated era was preceded by a vacuum
dominated era.

Inflation ends on a spacelike hypersurface $V$ at $t = t_{\rm
reheat}$. At this time, all the entropy in the universe is produced
through reheating.  Both before and after reheating, all spheres will
be anti-trapped except in a small neighbourhood of $r=0$ (or small
neighbourhoods of the poles of the $S^3$ spacelike slice in a closed
universe), of the size of the apparent horizon.  Therefore the
spacelike projection theorem does not apply to any but the smallest of
the surfaces $B \subset V$.  The other spheres may be exponentially
large, but the covariant conjecture does not relate their area to the
enclosed entropy.  The size and total entropy of the reheating
hypersurface $V$ is thus irrelevant.

Outside the apparent horizon, entropy/area comparisons can only be
done on the light-sheets specified in the conjecture.  Indeed, the
past-directed light-sheets of anti-trapped surfaces of the
post-inflationary universe do intersect $V$.  Since there is virtually
no entropy during inflation, we can consider the light-sheets to be
truncated by the reheating surface.  Because inflation cannot produce
more than one unit of entropy per Planck volume, the bound will be
satisfied by the same arguments that were given above for universes
with a big bang.

\subsection{Normal surfaces}
\label{sec-normal}

Spatial regions enclosed by normal surfaces will turn out, in a sense
specified below, to be analogues to ``Bekenstein systems'' (see
Sec.~\ref{sec-bekbound}).  In order to understand this, we must
establish a few properties of Bekenstein systems.  \label{bekdefs} For
quick reference, we will call the first bound, Eq.~(\ref{eq-bek-em}),
{\em Bekenstein's entropy/mass bound}, and the second bound,
Eq.~(\ref{eq-bek-ea}), {\em Bekenstein's entropy/area bound}.  For a
spherical Bekenstein system of a given radius and mass, the
entropy/mass bound is always at least as tight as the entropy/area
bound.  This because a Bekenstein system must be gravitationally
stable ($M \leq R/2$), which implies $2 \pi R M \leq \pi R^2 = A/4$.
A spherical system that saturates Bekenstein's entropy/area bound will
be called a {\em saturated Bekenstein system}.  The considerations
above imply that this system will also saturate the entropy/mass
bound.  In semi-classical gravity, a black hole, viewed from the
outside, is an example of a saturated Bekenstein system; but for many
purposes it is simpler to think of an ordinary, maximally enthropic,
spherical thermodynamic system just on the verge of gravitational
collapse.  A system that saturates the entropy/mass bound but not the
entropy/area bound will be called a {\em mass-saturated Bekenstein
system}.  If we find a spherical thermodynamic system which saturates
the entropy/area bound but does not saturate the entropy/mass bound,
we must conclude that it was not a Bekenstein system in the first
place.

The normal region contains a past- and a future-directed
light-sheet. Both of them are ingoing, i.e. they are directed {\em
towards the center} of the region at $\chi=0$ (or $\chi=\pi$ for the
normal region near the opposite pole in closed universes).  This is
crucial.  If outgoing light-sheets existed even as $\chi \rightarrow
0$, the area would become arbitrarily small while the entropy remained
finite.  This is one of the reasons why the Fischler-Susskind proposal
does not apply to closed universes (see Sec.~\ref{sec-fs}).  Except
for this constraint, the past-directed light-sheet coincides with the
light-cones used by Fischler and Susskind.  The entropy/area bound has
been shown to hold on such surfaces~\cite{FisSus98,BakRey99,KalLin99}.
We will therefore consider only the future-directed light-sheet.

The future-directed light-sheet covers the same comoving volume as the
past light-sheet.  Therefore the covariant entropy bound will be
satisfied on it if the evolution is adiabatic.  But we certainly must
allow the possibility that additional entropy is produced.  Consider,
for example, the outermost surface on which future-directed
light-sheets are still allowed, a sphere $B$ on the apparent horizon.
Suppose that an overfunded group of experimental cosmologists within
the apparent horizon are bent on breaking the entropy bound.  They
must try to produce as much entropy as possible before the matter
passes through the future-directed light-sheet $L$ of $B$.  What is
their best strategy?

Note that they cannot collect any mass from outside $B$, because $L$
is a null hypersurface bounded by $B$ and the dominant energy
condition holds, preventing spacelike energy flow.  The most enthropic
system is a saturated Bekenstein system, so they should convert all
the matter into such systems.  (As discussed above, saturated
Bekenstein systems are just on the verge of gravitational collapse and
contain the same amount of entropy as a black hole of the same mass
and radius.  By using ordinary thermodynamic systems instead of black
holes, one ensures that the light-sheet actually permeates the systems
completely and samples all the entropy, rather than being truncated by
the black hole singularity; see Secs.~\ref{sec-twoballs} and
\ref{sec-collapse} for a discussion of these issues.)

If all matter is condensed into several small highly enthropic
systems, they will be widely separated, i.e. surrounded by empty
regions of space which are large, flat, and static compared to the
length scale of any individual system.  By the dominant energy
condition, no negative energy is present.  Therefore conditions given
in Sec.~\ref{sec-bekbound} are fulfilled.  We are thus justified in
applying Bekenstein's bounds to the systems, and we should use the
entropy/mass bound, Eq.~(\ref{eq-bek-em}), because it is tighter.  In
order to create the maximum amount of entropy, however, it is best to
put the matter into a few large saturated Bekenstein systems, rather
than many small ones.  Therefore we should take the limit of a
Bekenstein system, as large as the apparent horizon and containing the
entire mass within it.  Of course the question arises whether the
calculation remains consistent in this limit, both in its treatment of
the interior as a Bekenstein system, and in its evaluation of the
mass.  We will show below, purely from the point of view of
Bekenstein's bound, that the interior of the apparent horizon is in
fact the largest system for which Bekenstein's conditions can be
considered to hold; for larger systems, inconsistencies arise.  In
Sec.~\ref{sec-others}, we will use the spacelike projection theorem to
arrive at the same conclusion within the framework of the covariant
entropy conjecture.

We have cautioned in Sec.~\ref{sec-cec} against using an entropy/mass
bound in general space-times, because there is no concept of local
energy density.  In this case, however, the mass is certainly well
defined before we take the limit of a single Bekenstein system,
because the many saturated systems will be widely separated and can be
treated as immersed in asymptotically flat space.  In the limit of a
single system, we can follow Ref.~\cite{BakRey99} and treat the
interior of the apparent horizon as part of an oversized spherical
star.  (We are thus pretending that somewhere beyond the apparent
horizon, the space-time may become asymptotically flat and empty.
This is not an inconsistent assumption as long as Bekenstein's
conditions are satisfied; we show below that this is indeed the case.
Related ideas, not referring specifically to the apparent horizon,
underly some proposals for cosmological entropy bounds given in
Refs.~\cite{EasLow99,Ven99,KalLin99}.)  Then we can apply the
usual mass definition for spherically symmetric systems~\cite{MTW} to
the interior region.

The circumferential radius of our system is the apparent horizon
radius, and is given by Eq.~(\ref{eq-ah-rho}):
\begin{equation}
r_{\rm AH} = \sqrt{\frac{3}{8 \pi \rho}}.
\label{eq-ah-rho2}
\end{equation}
The mass inside the apparent horizon is given by~\cite{MTW}
\begin{equation}
M(r_{\rm AH}) = \int_0^{r_{\rm AH}} 4\pi r^2 \rho dr = \frac{4\pi}{3}
  r_{\rm AH}^3 \rho.
\end{equation}
This yields $r_{\rm AH} = 2 M(r_{\rm AH})$.  By Bekenstein's
entropy/mass bound, Eq.~(\ref{eq-bek-em}), the entropy cannot exceed
$2\pi M(r_{\rm AH})\, r_{\rm AH}$.  Thus we find
\begin{equation}
S_{\rm max} = \pi r_{\rm AH}^2.
\end{equation}
This is exactly a quarter of the area of the apparent horizon.

Eq.~(\ref{eq-ah-rho}) follows from the Friedmann equation (which
involves only the density but not the pressure), and from
Eq.~(\ref{eq-ah-hubble}), which is a geometric property of the FRW
metrics.  Thus the calculation holds independently of the equation of
state.  We have not dropped any factors of order one, and attained
exactly the saturation of the bound.  This would not be the case for
the Hubble horizon or the particle horizon.  We conclude that the
entropy on the future-directed light-sheet $L$ will not exceed a
quarter of the area of the boundary $B$.  The covariant entropy bound
may be saturated on the apparent horizon, but it will not be violated,
no matter how hard we try to produce entropy.

The property $r = 2M(r)$ is special to the apparent horizon.  It
suggests that we should consider the interior of the apparent horizon
to be the largest region with non-dominant self-gravity, and thus the
largest system to which Bekenstein's bound can be applied.  This
statement can be made more precise.  The property $r_{\rm AH} =
2M(r_{\rm AH})$ means that if we treat the interior as a Bekenstein
system, it can be saturated, not just mass-saturated (see
page~\pageref{bekdefs} for definitions).  If we chose a smaller
surface $r_X < r_{\rm AH}$, the enclosed mass would be less than half
of the radius, $2M < r_X$.  In this case, (area-)saturation would not
be possible, but only mass-saturation.  This is because the
entropy/mass bound yields $2\pi M r_X < \pi r_X^2 = A/4$.  We conclude
that the covariant entropy bound, which uses area, cannot be saturated
on such surfaces.  On the other hand, surfaces outside the apparent
horizon, $r_X > r_{\rm AH}$, possess no future-directed light-sheet to
which we could apply the covariant bound.  We find this reflected in
the property $2M > r_X$ for such surfaces.  It implies that one could
build a mass-saturated spherical system which breaks Bekenstein's
entropy/area bound: $2 \pi M r_X > \pi r_X^2$.  As we discussed at the
beginning of this subsection, this indicates a breakdown of the
treatment of the enclosed region as a Bekenstein system.  It follows
that the apparent horizon is the largest sphere whose interior may be
treated as a Bekenstein system.  For larger systems, Bekenstein's
assumptions would not be self-consistent.

From the point of view of the covariant entropy conjecture, there are
well-defined sufficient conditions for the treatment of a spatial
region as a Bekenstein system, namely those spelled out in the
spacelike projection theorem (Sec.~\ref{sec-spt}).  This will be
applied to cosmology in the Sec.~\ref{sec-coscol}.  We will show
formally that the region inside the apparent horizon is indeed the
largest region for which Bekenstein's entropy/area bound may be
guaranteed to hold, if certain additional conditions are met.

\sect{Cosmological entropy bounds}
\label{sec-proposals}

\subsection{A cosmological corollary}
\label{sec-coscol}

In Sec.~\ref{sec-normal}, we tested the covariant entropy conjecture
on future-directed light-sheets in normal regions, by assuming maximal
entropy production in the interior spatial region.  We found that the
bound may be saturated, but not violated.  We will now switch
viewpoints, assume that the covariant entropy conjecture is a correct
law, and derive an entropy bound for spatial regions in cosmology.

Normal regions offer an interesting application of the spacelike
projection theorem (Sec.~\ref{sec-spt}).  It tells us under which
conditions we can treat the interior of the apparent horizon as a
Bekenstein system.  Let $A$ be the area of a sphere $B$, on or inside
of the apparent horizon.  (``Inside'' has a natural meaning in normal
regions.)  Then the future-directed ingoing light-sheet $L$ exists.
Let us assume that it is complete, i.e., $B$ is its only boundary.
(This condition is fulfilled, e.g., for any radiation or dust
dominated FRW universe with no cosmological constant.)  Let $V$ be a
region inside of and bounded by $B$, on any spacelike hypersurface
containing $B$.  If no black holes are produced, $V$ will be in the
causal past of $L$, and the conditions for the spacelike projection
theorem are satisfied.  Therefore, the entropy on $V$ will not exceed
$A/4$.  In particular, we may choose $B$ to be on the apparent
horizon, and $V$ to be on the spacelike slice preferred by the
homogeneity of the FRW cosmologies.  We summarize these considerations
in the following corollary:

\paragraph{Cosmological Corollary}
{\em Let $V$ be a spatial region within the apparent horizon of an
observer. If the future-directed ingoing light-sheet $L$ of the
apparent horizon has no other boundaries, and if $V$ is entirely
contained in the causal past of $L$, then the entropy on $V$ cannot
exceed a quarter of the area of the apparent horizon.}\\

By definition, the spheres beyond the apparent horizon are
anti-trapped and possess no future-directed light-sheets.  Therefore
the spacelike projection theorem does not apply, and no statement
about the entropy enclosed in spatial volumes can be made.  (Of
course, we can still use their area to bound the entropy on their
light-sheets.)  Thus the covariant entropy conjecture singles out the
apparent horizon as a special surface.  It marks the largest surface
to which the spacelike projection theorem can possibly apply, and
hence the region inside it is the largest region one can hope to treat
as a Bekenstein system.  This conclusion agrees with the result
obtained in the previous subsection from a consistency analysis of
Bekenstein's assumptions for regions larger than the apparent horizon.

As an example, the conditions of this corollary are satisfied by the
apparent horizon of de Sitter space.  It coincides with the
cosmological horizon, at $r=(3/\Lambda)^{1/2}$.  Thus the entropy
within the cosmological horizon cannot exceed $3\pi/\Lambda$.  The
corollary will be further applied in Sec.~\ref{sec-others}.

The corollary tells us if and how Bekenstein's bound can be applied to
spatial regions in cosmological solutions.  It follows from, but is
not equivalent to, the covariant entropy bound.  Like the spacelike
projection theorem, the corollary is a statement of limited scope.  It
contains no information how to relate entropy to the area of trapped
or anti-trapped surfaces in the universe, and even for surfaces within
the apparent horizon, a ``spacelike'' bound applies only under certain
conditions.  Thus, the role of the corollary is to define the range of
validity of Bekenstein's entropy/area bound in cosmological solutions.
Precisely for this reason, the corollary is far less general than the
covariant conjecture, which associates at least two hypersurfaces with
any surface in any space-time, and bounds the entropy on those
hypersurfaces.


\subsection{The Fischler-Susskind bound}
\label{sec-fs}

Among the recently proposed cosmological entropy
bounds~\cite{FisSus98,EasLow99,Ven99,BakRey99,KalLin99,Bru99}, the
prescription of Fischler and Susskind (FS) is distinct in that it
attempts to relate entropy to every spherical surface in the universe,
namely the entropy on the past-directed ``ingoing'' null hypersurface.
The covariant entropy conjecture is very much in this spirit; but we
have changed the prescription from ``past-ingoing'' to a general
selection rule determining at least two light-sheets on which entropy
can be compared to area.  The FS hypersurface will often be one of
them, but not always.  The limitations of the FS proposal can be
understood in terms of this selection rule.

Consider the entropy on the null hypersurface formed by the particle
horizon, emanating from the South pole in a closed, adiabatic,
dust-dominated FRW universe~\cite{FisSus98}.  The solution is given by
\begin{equation}
a(\eta) = \frac{a_{\rm max}}{2} ( 1 - \cos \eta ),~~~
t(\eta) = \frac{a_{\rm max}}{2} ( \eta - \sin \eta ).
\end{equation}
The entropy may not exceed one per Planck volume at the Planck time,
$\eta_{\rm Pl} \sim a_{\rm max}^{-1/3}$.  Therefore the total entropy
of the universe will not be larger than $S_{\rm max} \sim a(\eta_{\rm
Pl})^3 \sim a_{\rm max}$.

\begin{figure}[htb!]
  \hspace{.18\textwidth} \vbox{\epsfxsize=.64\textwidth
  \epsfbox{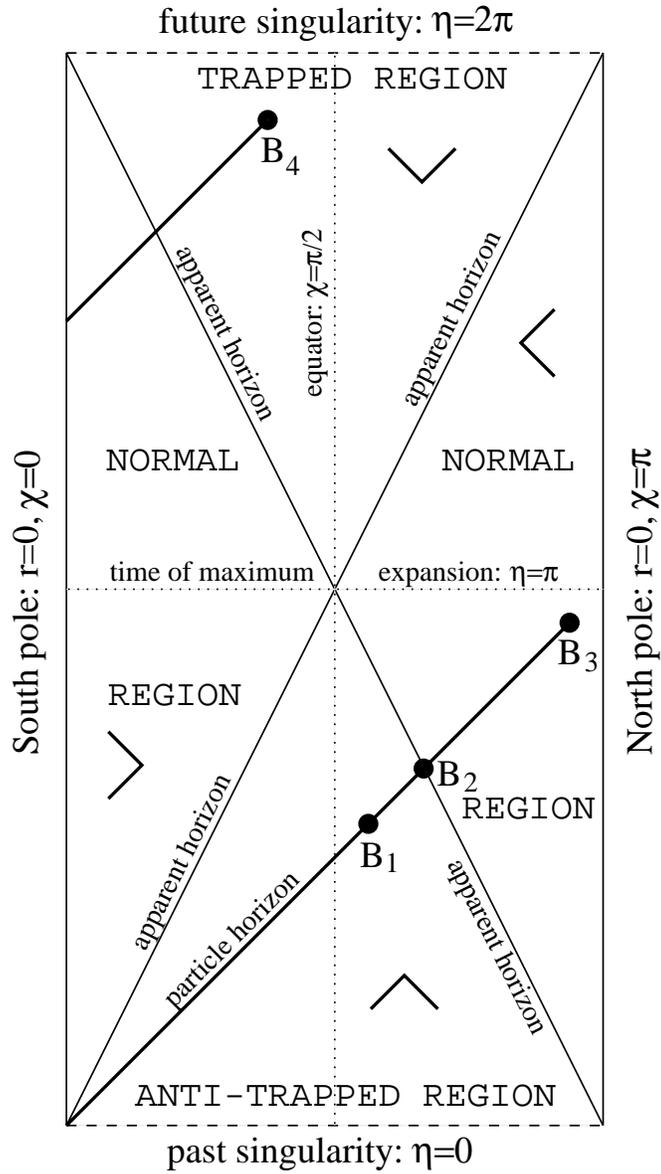}}
\caption%
{\small\sl Penrose diagram for a closed FRW universe dominated by
dust.  The horizontal lines correspond to $S^3$ spacelike sections.
Every point represents a two-sphere.  The apparent horizons divide the
space-time into four regions.  The directions of light-sheets in each
region are indicated by small wedges.}
\label{fig-closeddust}
\end{figure}

The particle horizon has $\eta = \chi$, while the apparent horizon is
given by $\eta = 2\chi$ (see Fig.~\ref{fig-closeddust}).
Thus the particle horizon will initially be outside the apparent
horizon, in an anti-trapped region.  Any surfaces met by the particle
horizon in this region (such as $B_1$) possess a past-directed
light-sheet that coincides with the particle horizon.  Therefore the
FS bound and the covariant bound should both be satisfied.  Before the
time of maximum expansion ($\eta = \pi$), the particle horizon reaches
the sphere $B_2$ at $\eta = \chi = 2\pi/3$.  Let us verify explicitly
that the bounds are still satisfied there.  At this time, the particle
horizon covers two-thirds of the total entropy of the universe, which
will be of order $a_{\rm max}$. Its area, however, is $4 \pi (3/4)^3
a_{\rm max}^2$, which is much larger.

For $\eta = \chi > 2\pi/3$ the particle horizon ends on normal, rather
than anti-trapped spheres.  It has entered a normal region surrounding
the North pole and bounded by a different apparent horizon, $\eta =
2(\pi-\chi)$.  The surfaces in this region contain a future- and a
past-directed light-sheet, both pointing towards the North pole.  The
particle horizon, which goes to the South pole, is now one of the
``forbidden'' families of light-rays (Sec.~\ref{sec-recipe}).
According to the covariant bound, the entropy contained on it has
nothing to do with the surface that bounds it.  Indeed, its area
approaches zero when it encompasses nearly the entire entropy of the
universe ($B_3$).  The FS bound cannot be applied here.  It would
continue to compare entropy and area of the particle horizon, and
would be violated~\cite{FisSus98}.

In Sec.~\ref{sec-cec}, the selection rule was motivated by the
requirement that one should compare the area of a surface only to the
entropy that is, in some sense, inside of it, not outside.  This
consideration is paying off here.  The spheres very close to the North
pole, like any other $S^2$ on an $S^3$, enclose two regions.  Because
the region including the North pole is much smaller than the region
including the South pole, one would like to call the former the
``inside'' and the latter the ``outside.''  The selection rule turns
this intuitive notion into a covariant definition, which takes into
account not only the shape of space but also its dynamics.  Trapped
surfaces, for example, have light-sheets on both sides, but only
future-directed ones.  This makes sense because in a collapsing
region, loosely speaking, the direction in which surfaces are getting
smaller is the future.

Returning to the example of a closed, adiabatic, dust-dominated FRW
universe, consider a surface $B_4$ in the trapped region, near the
equator of the $S^3$ spacelike surfaces (see
Fig.~\ref{fig-closeddust}).  By choosing $B_4$ to be very close to the
future singularity, its area can be made arbitrarily small.  The FS
hypersurface would go to the past and would pick up nearly half of the
total entropy, so the FS bound cannot be used in this region.  The
covariant bound remains valid, because it applies only to the
future-directed null hypersurfaces, which are soon truncated by the
singularity.

\subsection{Other cosmological entropy bounds}
\label{sec-others}

Other proposals for cosmological entropy
bounds~\cite{EasLow99,Ven99,BakRey99,KalLin99,Bru99} are based on the
idea of defining a limited spatial region to which Bekenstein's bound
can be applied.%
\footnote{We should point out that the first application of any
entropy bound to cosmology was by Bekenstein, in Ref.~\cite{Bek89}.}
The definitions refer variously to the Hubble horizon
(or to a region of size $\sim H^{-1} =
a/\dot{a}$)~\cite{EasLow99,Ven99,KalLin99,Bru99}, a region of size
$\sim t$~\cite{KalLin99}, and, remarkably, the apparent
horizon~\cite{BakRey99}.  In its simplified version, the
Fischler-Susskind proposal can be included in this class, as referring
to the region within the particle horizon~\cite{FisSus98}.

The prescriptions do not aim to relate entropy to any surfaces larger
than the specified ken.  Also, most do not claim validity during the
collapsing era of a closed universe, and none can be applied in
arbitrary gravitationally collapsing systems.  The covariant entropy
bound differs from this approach in that it attempts generality: It
associates hypersurfaces with {\em any} surface in {\em any}
space-time and bounds the entropy contained on those hypersurfaces.

The proposed cosmological bounds are very useful for estimating the
maximal entropy in limited regions of cosmological solutions.  Even if
a horizon other than the apparent horizon is used, one may still
obtain correct results at least within factors of order one.  In order
to avoid pitfalls, however, they must be used carefully (as many
authors have stressed).  The corollary derived in
Sec.~\ref{sec-coscol} contains precise conditions determining whether,
and to which regions of the universe, Bekenstein's bound can be
applied.

Consider, for example, the bound of Bak and Rey~\cite{BakRey99}, which
refers explicitly to the spatial region within the apparent horizon.
One might be tempted to consider this as a special case of the
covariant entropy bound, in the sense of the corollary derived in
Sec.~\ref{sec-coscol}.  Then it should always be valid.  But Kaloper
and Linde~\cite{KalLin99} have shown that this bound is exceeded in
flat, adiabatic FRW universes with an arbitrarily small but
non-vanishing negative cosmological constant.  So what has gone wrong?

\begin{figure}[htb!]
  \hspace{.12\textwidth} \vbox{\epsfxsize=.76\textwidth
  \epsfbox{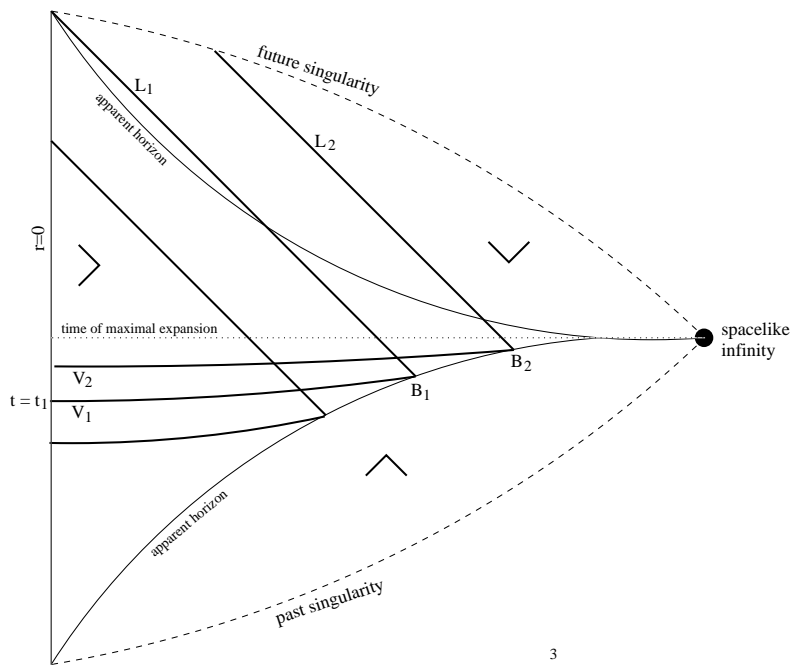}}
\caption%
{\small\sl Penrose diagram for a flat universe with matter and a
negative cosmological constant.  Two apparent horizons divide the
space-time into an anti-trapped, a trapped, and a normal region.  The
wedges show the light-sheet directions.  The future-directed ingoing
light-sheet $L_1$ of the surface $B_1$ is complete. By the spacelike
projection theorem (Sec.~\ref{sec-spt}), the entropy on the spatial
region $V_1$ is bounded by the area of $B_1$.  An apparent horizon
surface $B_2$ at a later time, however, does not possess a complete
future-directed ingoing light-sheet: $L_2$ is truncated by the future
singularity.  Therefore the area of $B_2$ bounds only the entropy on
its three light-sheets, not on its spatial interior $V_2$.}
\label{fig-bakrey}
\end{figure}

As the Penrose diagram for this space-time (Fig.~\ref{fig-bakrey})
shows, the universe starts out much like an ordinary flat FRW
universe.  As the normal matter is diluted, it eventually becomes
dominated by the negative cosmological constant.  This slows down the
expansion of the universe so much that it starts to recollapse.  The
evolution is symmetric about the turn-around time.  Eventually, matter
dominates again and the universe ends in a big crunch.  As the
turnaround time is approached, the apparent horizon moves out to
spatial infinity~\cite{KalLin99}, and the enclosed volume grows
without bound.  The entropy density is constant, and the area grows
more slowly than the volume.  Thus the entropy eventually exceeds the
Bak-Rey bound.

But the cosmological corollary (Sec.~\ref{sec-coscol}) states that one
can use the interior of the apparent horizon for an entropy/area bound
only if the future-directed ingoing light-sheet is complete.  As
Fig.~\ref{fig-bakrey} shows, this ceases to be the case before the
turnaround time.  After the time $t_1$, where $t_1 < t_{\rm
turnaround}$, the future-directed light-sheet of the apparent horizon
will have another boundary, namely on the future singularity of the
space-time.  Thus the conditions for the spacelike projection theorem,
and the corollary it implies, are no longer met.  The entropy in the
spatial interior of the apparent horizon~\cite{BakRey99} will not be
bounded by its area in this region.

The covariant entropy conjecture states that the entropy on the
future-directed ingoing light-sheet, as well as the two past-directed
light-sheets of the apparent horizon will each be less than $A/4$.
Because the future-directed light-sheet is truncated by the future
singularity, and the past-directed light-sheets are truncated by the
past singularity, the comoving volume swept out by any of these
light-sheet grows only like the area as one moves further along the
apparent horizon.  Thus there is no contradiction with the covariant
bound.

Neither is there any contradiction with the calculation performed in
Sec.~\ref{sec-normal}, which concluded that Bekenstein's bound can be
applied to the region within the apparent horizon.  This calculation
was done within the framework of Bekenstein's conditions, and thus
explicitly assumed the positivity of energy, which is violated here.
As we pointed out in Sec.~\ref{sec-bekbound}, Bekenstein's bound
cannot be applied to regions containing a negative energy component,
such as a negative cosmological constant.  This indicates a striking
difference to the covariant bound.  While we have formally assumed the
positivity of energy as a condition for the covariant entropy bound,
we have argued in Sec.~\ref{sec-cec} that the validity of the bound
may well extend to space-times with a negative cosmological constant.
We find ourselves encouraged by this example.

\sect{Testing the conjecture in gravitational collapse}
\label{sec-collapse}

In Sec.~\ref{sec-cosmology} we tested the covariant entropy conjecture
for anti-trapped and normal surfaces in cosmological space-times.  We
found that even the most non-adiabatic processes can only saturate,
but not violate, the bound.  We now turn to trapped surfaces, which
occur not only in collapsing universes, but arise generally during
gravitational collapse and inside black holes.

Like anti-trapped regions, trapped regions are manifestly dominated by
self-gravity, and Bekenstein's bound will be of little help.  The
covariant entropy bound must be justified by other considerations.
This is a more subtle problem for trapped, than for anti-trapped
surfaces.  It was reasonable to require (Sec.~\ref{sec-antitrapped})
that the entropy one Planck time away from a past singularity cannot
exceed one per Planck volume; this merely amounts to a sensible
specification of initial conditions.  Because of the rapid expansion,
the entropy/area ratio decreases at later times, and a situation in
which the covariant bound would be exceeded does not arise.  But near
future singularities, one cannot use the time-reverse of this argument
to ``retrodict'' that some experimental setup was impossible to start
with.  Initial conditions are set in the past, not in the future.

At first sight it seems obvious that the covariant entropy bound, like
any entropy/area bound, will be violated in trapped regions.  Consider
a saturated Bekenstein system of area $A_0$, in which gravitational
collapse is induced.  The system will shrink, but by the second law,
the entropy will not decrease.  A short time after the beginning of
the collapse, the surface of the system will have an area $A_1<A_0$.
Because the surface is trapped, the past-directed ingoing light-rays
have positive expansion and cannot be considered.  But there will be a
future-directed ingoing light-sheet.  If this light-sheet penetrated
the entire system, it would contain an entropy of $S \geq A_0/4 >
A_1/4$ and the bound would be exceeded.  There are a number of
effects, however, which constrain the extent to which the light-sheet
can sample entropy.  We will discuss them qualitatively, before
turning to a quantitative test in Sec.~\ref{sec-miracle}.

\subsection{Light-sheet penetration into collapsing systems}
\label{sec-penetration}

When the boundary of the system is a suffiently small proper time away
from the future singularity, the light-sheet will not intersect the
whole system, because it will be truncated by the singularity (or a
surface where the Planck density is reached), much like the past
light-sheet of spheres larger than the particle horizon in a flat FRW
universe.  {\em Truncation} is a basic limitation~\cite{EasLow99}, but
additional constraints will be needed.

\begin{figure}[htb!]
  \hspace{.15\textwidth} \vbox{\epsfxsize=.7\textwidth
  \epsfbox{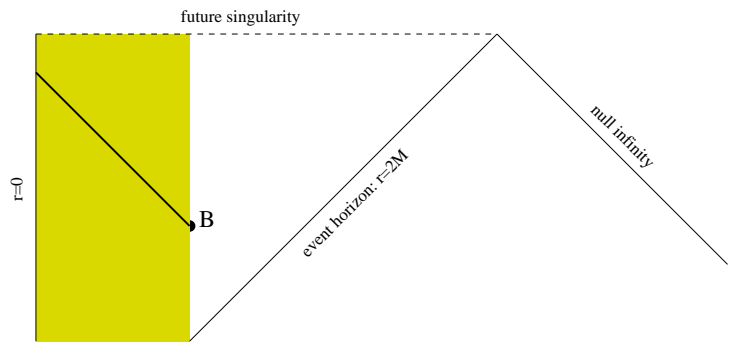}}
\caption%
{\small\sl Penrose diagram for the Oppenheimer-Snyder collapse of a
dust star.  The interior of the star is described by a portion of the
closed dust-dominated FRW solution.  Correspondingly, it is
represented by the top left
quarter of the Penrose diagram, Fig.~\ref{fig-closeddust}. The
exterior is a portion of the Schwarzschild solution.  Light-rays
emitted from the surface of the star when it is inside its
gravitational radius can traverse the interior entirely.  This
is not a violation of the covariant entropy bound because a collapsing
dust star does not have much entropy.  For highly enthopic collapsing
objects, a number of effects must be taken into account which protect
the conjecture (see text).}
\label{fig-oscollapse}
\end{figure}

Consider the Oppenheimer-Snyder collapse of a dust ball~\cite{MTW},
commencing from a momentarily static state with $R=2M$, as shown in
Fig.~\ref{fig-oscollapse}.
A future-directed ingoing light-sheet, starting at the surface $B$ at
a sufficiently early time but inside the event horizon, can easily
traverse the ball before the singularity (or the Planck density, at $R
\sim M^{1/3}$) is reached.  But this light-sheet would endanger the
bound only if the system collapsed from a state in which it nearly
saturated Bekenstein's bound.  So how much entropy does the dust star
actually contain?  Strictly, the Oppenheimer-Snyder solution describes
a dust ball at zero temperature.  Since it also must be exactly
homogeneous, it contains not even the usual positional entropy equal
to the particle number.  Thus the entropy is zero.  In order to
introduce a sizable amount of entropy, we have to violate the
conditions under which the solution is valid: homogeneity and zero
temperature.  This collapse will be described by a different solution,
for which a detailed calculation would have to be done to determine
the penetration depth of light-sheets.

By definition, highly enthropic systems undergoing gravitational
collapse are very irregular internally and contain strong small scale
density perturbations.  This will make the collapse inhomogeneous,
with some regions reaching the singularity after a shorter proper time
than other regions.  One might call this effect {\em Local
Gravitational Collapse}.  A saturated Bekenstein system is globally
just on the verge of gravitational collapse: $R = 2M$.  But as it
contracts, individual parts of the system, of size $\Delta R < R$ will
become gravitationally unstable: $\Delta R \leq 2 \Delta M$.
Particles in an overdense region will reach a singularity after a
proper time of order $\Delta R^2/ \Delta M$, which is shorter than the
remaining lifetime of average regions, $\sim R^2/M$.  This makes it
more difficult for a light-sheet to penetrate the system completely,
unless it originates near the beginning of the collapse, when the area
is still large.

The internal irregularity of highly enthropic systems also enhances
the effect of {\em Percolation}, discussed in Sec.~\ref{sec-twoballs}.
Inhomogeneities that break spherical symmetry will deflect the rays in
the light-sheet and cause dents in their spherical cross-sections.
Such dents will develop into ``angular'' caustics.  At any caustic,
the light-sheet ends, because the expansion becomes positive
(Sec.~\ref{sec-caustics}).  Any light-rays that do not end on caustics
will follow an irregular path through the object similar to random
walk.  Since they waste affine time on covering angular directions,
they may not proceed far into the object before the singularity is
reached.  Thus it may well be impossible for a light-sheet to
penetrate through a collapsing, highly entropic system far enough to
sample excessive entropy.

The quantitative investigation of the formation of angular caustics on
light-sheets penetrating collapsing, highly enthropic systems lies
beyond the scope of this paper.  A strong, quantitative case for the
validity of the bound may still be made by eliminating the percolation
effect.  We will consider a system containing only radial modes.  This
system is spherically symmetric even microscopically, and cannot
deflect light-rays into angular directions.  With no constraints on
mass and size, it can contain arbitrary amounts of entropy, but cannot
lead to angular caustics on the light-sheet.

\subsection{A quantitative test}
\label{sec-miracle}

\begin{figure}[htb!]
  \hspace{.2\textwidth} \vbox{\epsfxsize=.6\textwidth
  \epsfbox{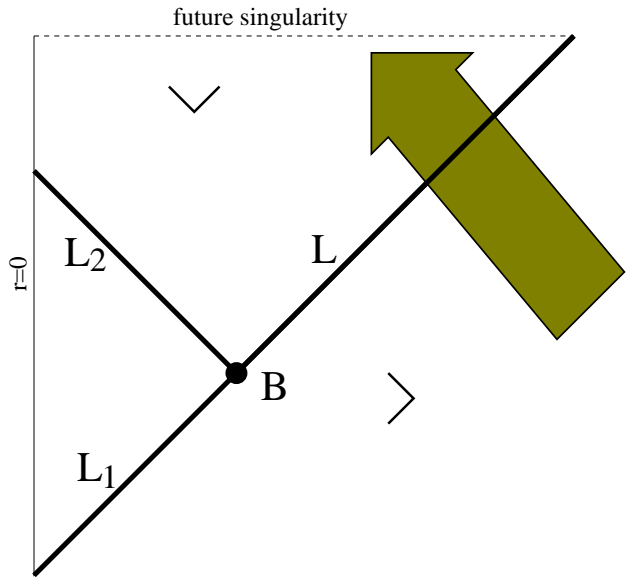}}
\caption%
{\small\sl Penrose diagram for a black hole.  $B$ is a surface on the
apparent horizon and has three light-sheets: $L_1$, $L_2$, and $L$. In
an attempt to violate the covariant entropy conjecture, we collapse a
massive shell around the black hole.  We find, however, that one cannot
squeeze more entropy through $L$ than a quarter of the area of $B$.
Otherwise, $L$ reaches the singularity before passing through the
infalling shell entirely.}
\label{fig-shellcollapse}
\end{figure}

Consider a Schwarzschild black hole of horizon size $r_0$ (see
Fig.~\ref{fig-shellcollapse}).
Let $B$ be a sphere on the apparent
horizon at some given time; thus $B$ has area
\begin{equation}
A = 4 \pi r_0^2.
\label{eq-aharea}
\end{equation}
The surface $B$ is marginally outer trapped and possesses three
light-sheets.  The past-directed ingoing light-sheet, $L_1$ counts the
entropy $S_{\rm in}$ that went into the black hole.  The generalized
second law of thermodynamics guarantees that the entropy conjecture
will hold on this light-sheet, since
\begin{equation}
A/4 = S_{\rm bh} \geq S_{\rm in}.
\end{equation}
The future-directed ingoing light-sheet, $L_2$, may intersect
some or all of the collapsing object that formed the black hole.  The
future-directed outgoing light-sheet $L$ intersects objects falling
into the black hole at a later time.  The validity of the conjecture
for the future-directed light-sheets was supported by qualitative
arguments in the previous subsection.  For a quantitative test, we
shall now try to send excessive entropy through $L$.

$L$ has zero expansion and defines the black hole horizon as long as
no additional matter falls in.  When $L$ does encounter matter, the
expansion becomes negative and $L$ collapses to $r=0$ within a finite
affine time.  Our strategy will be to use an infalling shell of matter
to squeeze as much entropy as possible across $L$ before the
light-sheet ends on a caustic or a singularity.  We shall not have any
scruples about making the mass of the shell extremely large compared
to the mass of the black hole.  By preparing the shell far outside the
black hole and letting it collapse, we can thus transport an arbitrary
amount of entropy to $r_0$.  This means, of course, that the shell may
be well inside its own Schwarzschild radius by the time it reaches
$L$; but by the second law, this cannot reduce the entropy it carries.
What we must ensure, however, is that $L$ actually penetrates the
entire shell and samples all of its entropy.  We will now show that
this requirement keeps the shell entropy within the conjectured bound.

Let $\theta$ be the expansion of the null generators of $L$.  Its rate
of change is given by Raychauduri's equation, Eq.~(\ref{eq-raych}),
from which we obtain the inequality
\begin{equation}
\frac{d\theta}{d\lambda} \leq
     - \frac{1}{2} \theta^2 - 8\pi T_{ab} k^a k^b.
\label{eq-raych3}
\end{equation}
The first term on the right hand side is non-positive, and the
dominant energy condition is sufficient to ensure that the second term
is also non-positive.  Initially, $\theta=0$, since $L$ is an apparent
horizon.  Now consider a shell of matter, of mass $M$, crossing the
light-sheet $L$.  We would like to make the shell as wide as possible
in order to store a lot of entropy.  But we also must keep it
sufficiently thin, so that the light-sheet does not collapse due to
the first term in Eq.~(\ref{eq-raych3}), before the shell has
completely crossed it.

The maximum width of the shell can be easily estimated.  Consider an
infinitely thin shell of mass $M$ falling towards the black hole at
$r_0$.  Outside the shell, the metric will be given by a Schwarzschild
black hole of mass
\begin{equation}
\tilde{M} = M + \frac{r_0}{2},
\label{eq-mtilde}
\end{equation}
by Birkhoff's theorem.  Once the shell has crossed the light-sheet $L$
at $r_0$, the null generators of $L$ will be moving in a Schwarzschild
interior of mass $\tilde{M}$.  Therefore this meeting occurs at proper
time
\begin{equation}
\Delta\tau_{\rm dead} = r_0 + 2 \tilde{M} \ln \left(1 -
\frac{r_0}{2\tilde{M}} \right) \approx \frac{r_0^2}{4\tilde{M}}
\end{equation}
before the generators reach a singularity.  We may approximate a shell
of finite proper thickness $w$ by an infinitely thin shell of the same
mass, located $w/2$ from either side.  We are thus pretending that the
shell mass contributes to the second term of the Raychauduri equation
all at once, at the moment when half of the shell has already passed
through $L$.  Since we require that the light-sheet penetrates the
shell entirely, we must be sure that the other half of the shell
passes through $L$ before the light-sheet hits the singularity.  This
requires $w/2 \leq \Delta\tau_{\rm dead}$, whence
\begin{equation}
w_{\rm max} \approx r_0^2/2\tilde{M}.
\label{eq-wmax}
\end{equation}
The maximum width of the shell is thus inversely proportional to its
mass, and is always less than $r_0$.

The next step will be to calculate the maximum entropy of a spherical
shell of mass $M$ and width $w$.  We will build the shell at $r=R$,
far away from the black hole and outside its own gravitational radius;
i.e., $R$ is much greater than any of the quantities $w$, $M$, and
$r_0$.  Thus Bekenstein's bound can be used to estimate the entropy.
Then we will let the shell collapse into the black hole.  Since
$w_{\rm max}<r_0/2$ by Eq.~(\ref{eq-wmax}), the width of the shell
will be smaller than the curvature of space during the entire time of
the collapse, and we can take it to remain constant.  This enables us
to neglect effects of local gravitational collapse (which, if
included, constrain the setup further; see
Sec.~\ref{sec-penetration}).  In order to exclude effects of angular
caustics, which are difficult to deal with quantitatively, we must
specify that the shell contain only spherically symmetric
micro-states, i.e., micro-states living in the radial, not the angular
directions.  We will now estimate the maximum entropy of the shell.

The shell can be split into a large number $n=R^2/w^2$ of roughly
cubic boxes of volume $w^3$ and mass $M/n$, separated by impenetrable
radial walls.  By Eq.~(\ref{eq-bek-em}), the maximum entropy of an
isolated, single box is the largest dimension times the mass:
\begin{equation}
S_{\rm box} = 2 \pi w \frac{M}{n}.
\end{equation}
Since all states are restricted to be radial, no new states are added
by removing the wall between two adjacent boxes.  Thus the entropy of
two boxes will simply be twice the entropy of one box.  By repeating
this argument, we can remove all the separating walls.  Therefore the
shell has a maximum entropy of
\begin{equation}
S_{\rm shell} = 2 \pi w M.
\end{equation}
The width $w$ is restricted by Eq.~(\ref{eq-wmax}).  One has
$M/\tilde{M} \leq 1$ by Eq.~(\ref{eq-mtilde}), even if the limit $M
\rightarrow \infty$ is permitted.  Thus the shell entropy cannot
exceed $\pi r_0^2$, which by Eq.~(\ref{eq-aharea}) is a quarter of the
area bounding the light-sheet:
\begin{equation}
S_{\rm shell} \leq \frac{A}{4}.
\end{equation}

We conclude that $A/4$ is the maximum amount of entropy one can
transport through a future-directed outgoing light-sheet $L$ bounded
by a black hole apparent horizon of area $A$, using an exactly
spherically symmetric shell of matter.  We take this as strong
evidence in favor of the entropy bound we propose.

\sect{Conclusions}
\label{sec-conclusions}

Bekenstein has shown that the entropy of a thermodynamic system with
limited self-gravity is bounded by its area.  By demanding general
coordinate invariance and constructing a selection rule, we arrived at
a bound on the entropy present on null hypersurfaces in arbitrary
space-times.  We tested the conjecture on cosmological solutions and
inside gravitationally collapsing regions.  We found, under the most
adverse assumptions, that the bound can be saturated but not exceeded.
This evidence suggests that the covariant entropy conjecture may be a
universal law of physics.

But can the conjecture be proven?  The processes by which the bound is
protected appear to be rather subtle.  They differ according to the
physical situation studied, and they can involve combinations of
different effects more reminiscent of a conspiracy than of an elegant
mechanism (Sec.~\ref{sec-penetration}).  (This is quite in contrast to
Bekenstein's bound, which is protected by gravitational collapse; see
below.)  We have verified the bound in a wide class of space-times and
space-time regions, but from the perspective of general relativity,
the processes protecting the bound appear eclectic, and its success
remains mysterious.  This indicates that we may be looking at nature
in an artificial and complicated way when we describe it as a
$3+1$-dimensional space-time filled with matter.  If the covariant
bound is correct, we believe it must arise from a more fundamental
description of nature in an obvious way.  But this does not exclude
the possibility that it can be proven (in a complicated way) entirely
within general relativity.  The proof would have to combine the tools
used for establishing the laws of black hole mechanics~\cite{BarCar73}
with the formalism of the Hawking-Penrose singularity
theorems~\cite{HawEll}.

In the final paragraphs of Secs.~\ref{sec-recipe} and \ref{sec-spt} we
discussed the most important property of the conjecture: It is
manifestly time reversal invariant.  Therefore the second law of
thermodynamics, which underlies Bekenstein's bound, cannot be
responsible for the covariant bound, and one is forced to contemplate
the possibility of a different origin.

As a thermodynamic concept, entropy has a built-in arrow of time.  The
T-invariance of the covariant entropy conjecture can be understood
only if the bound is interpreted as a bound on the number of degrees
of freedom of the matter systems present on the light-sheets.  This
number is always at least as large as the thermodynamic entropy.  With
this statistical interpretation, T-invariance is natural.  However, we
never made any assumptions about the microscopic properties of matter,
which would limit the number of degrees of freedom present.  This
leaves no choice but to conclude that the number of degrees of freedom
in nature is {\em fundamentally} limited, as proposed in
Refs.~\cite{Tho93,Sus95}.

The idea that the world is effectively two-dimensional was put forth
by 't~Hooft~\cite{Tho93} and was further developed by
Susskind~\cite{Sus95}.  Based on Bekenstein's bound, the holographic
hypothesis was a bold leap.  One could argue, after all, that
Bekenstein's bound is not a fundamental limit on the number of degrees
of freedom, but a practical restriction on thermodynamic entropy.
There could be far more degrees of freedom in a system than its
surface area, but if too many of them were excited at the same time, a
black hole would form and the system would no longer satisfy
Bekenstein's conditions.  One may view Bekenstein's bound as arising
from this elegant gravitational collapse mechanism.  Bekenstein's law
applies only to Bekenstein systems; it works because a Bekenstein
system exceeding the bound ceases to be one.

The covariant entropy bound, on the other hand, applies even to
surfaces in collapsing regions, or on cosmological scales.  Its
generality, together with its T-invariance, force us to the conclusion
that a holographic principle underlies the description of nature.
Moreover, the bound leads naturally to a background-independent
formulation of the principle.  {\em The number of independent degrees
of freedom on any light-sheet of a surface $B$ cannot exceed a quarter
of the area of $B$.}

\section*{Acknowledgments}

I thank Gerard 't~Hooft, Nemanja Kaloper, Andrei Linde, and Lenny
Susskind for many extensive discussions.  This work has benefitted in
countless ways from their criticism and encouragement.  I am also
indebted to Jacob Bekenstein and Werner Israel for helpful
correspondence, and to Ted Jacobson for valuable comments on an
earlier version of this paper.


\begin{thebibliography}{10}

\bibitem{Bek81}
J.~D. Bekenstein: {\em A universal upper bound on the entropy to energy ratio
  for bounded systems\/}. Phys. Rev. D {\bf 23}, 287 (1981).

\bibitem{Bek72}
J.~D. Bekenstein: {\em Black holes and the second law\/}. Nuovo Cim. Lett. {\bf
  4}, 737 (1972).

\bibitem{Bek73}
J.~D. Bekenstein: {\em Black holes and entropy\/}. Phys. Rev. D {\bf 7}, 2333
  (1973).

\bibitem{Bek74}
J.~D. Bekenstein: {\em Generalized second law of thermodynamics in black hole
  physics\/}. Phys. Rev. D {\bf 9}, 3292 (1974).

\bibitem{Haw74}
S.~W. Hawking: {\em Particle creation by black holes\/}. Commun. Math. Phys.
  {\bf 43}, 199 (1974).

\bibitem{Bek94b}
J.~D. Bekenstein: {\em Do we understand black hole entropy?\/} gr-qc/9409015.

\bibitem{Pag76}
D.~N. Page: {\em Particle emission rates from a black hole. {II}. {M}assless
  particles from a rotating hole\/}. Phys. Rev. D {\bf 14}, 3260 (1976).

\bibitem{SchBek89}
M.~Schiffer and J.~D. Bekenstein: {\em Proof of the quantum bound on specific
  entropy for free fields\/}. Phys. Rev. D {\bf 39}, 1109 (1989).

\bibitem{FisSus98}
W.~Fischler and L.~Susskind: {\em Holography and cosmology\/}, hep-th/9806039.

\bibitem{Tho93}
G.~'t~Hooft: {\em Dimensional reduction in quantum gravity\/}, gr-qc/9310026.

\bibitem{Sus95}
L.~Susskind: {\em The world as a hologram\/}. J. Math. Phys. {\bf 36}, 6377
  (1995), hep-th/9409089.

\bibitem{EasLow99}
R.~Easther and D.~A. Lowe: {\em Holography, cosmology and the second law of
  thermodynamics\/}, hep-th/9902088.

\bibitem{Ven99}
G.~Veneziano: {\em Pre-bangian origin of our entropy and time arrow\/},
  hep-th/9902126.

\bibitem{BakRey99}
D.~Bak and S.-J. Rey: {\em Cosmic holography\/}, hep-th/9902173.

\bibitem{KalLin99}
N.~Kaloper and A.~Linde: {\em Cosmology vs. holography\/}, hep-th/9904120.

\bibitem{Bru99}
R.~Brustein: {\em The generalized second law of thermodynamics in cosmology\/},
  gr-qc/9904061.

\bibitem{CorJac96}
S.~Corley and T.~Jacobson: {\em Focusing and the holographic hypothesis\/}.
  Phys. Rev. D {\bf 53}, 6720 (1996), gr-qc/9602043.

\bibitem{HawEll}
S.~W. Hawking and G.~F.~R. Ellis: {\em The large scale stucture of
  space-time\/}. Cambridge University Press, Cambridge, England (1973).

\bibitem{Wald}
R.~M. Wald: {\em General Relativity\/}. The University of Chicago Press,
  Chicago (1984).

\bibitem{MTW}
C.~W. Misner, K.~S. Thorne and J.~A. Wheeler (eds.): {\em Gravitation\/}.
  Freeman, New York (1973).

\bibitem{Bek89}
J.~D. Bekenstein: {\em Is the cosmological singularity thermodynamically
  possible?\/} Int. J. Theor. Phys. {\bf 28}, 967 (1989).

\bibitem{BarCar73}
J.~M. Bardeen, B.~Carter and S.~W. Hawking: {\em The four laws of black hole
  mechanics\/}. Commun. Math. Phys. {\bf 31}, 161 (1973).

\end{thebibliography}

\end{document}